\documentstyle[12pt]{article}
\newlength{\dinwidth}
\newlength{\dinmargin}
\newlength{\extraspace}
\newlength{\extraspaces}
\newcommand{\be}{\begin{equation}
\addtolength{\abovedisplayskip}{\extraspaces}
\addtolength{\belowdisplayskip}{\extraspaces}
\addtolength{\abovedisplayshortskip}{\extraspace}
\addtolength{\belowdisplayshortskip}{\extraspace}}
\newcommand{\ee}{\end{equation}}
\newcommand{\bdm}{\begin{displaymath}
\addtolength{\abovedisplayskip}{\extraspaces}
\addtolength{\belowdisplayskip}{\extraspaces}
\addtolength{\abovedisplayshortskip}{\extraspace}
\addtolength{\belowdisplayshortskip}{\extraspace}}
\newcommand{\edm}{\end{displaymath}}
\renewcommand{\thefootnote}{\fnsymbol{footnote}}
\def\simlt{\mathrel{\lower2.5pt\vbox{\lineskip=0pt\baselineskip=0pt
           \hbox{$<$}\hbox{$\sim$}}}}
\newcommand{\SM}{Standard Model}
\newcommand{\la}{\lambda}
\newcommand{\La}{\Lambda}
\newcommand{\cD}{{\cal D}}
\newcommand{\pr}{Phys.\ Rev.\ }
\newcommand{\prp}{Phys.\ Rep.\ }
\newcommand{\prl}{Phys.\ Rev.\ Lett.\ }
\newcommand{\np}{Nucl.\ Phys.\ {\bf B}}
\newcommand{\pl}{Phys.\ Lett.\ {\bf B}}
\newcommand{\jsp}{J.\ Stat.\ Phys.\ }
\newcommand{\cmp}{Comm.\ Math.\ Phys.\ }
\newcommand{\zp}{Z. Phys.\ {\bf C}}
\newcommand{\ptp}{Prog.\ Theor.\ Phys.\ }
\newcommand{\ap}{Ann.\ Phys.\ }
\begin{document}
\begin{titlepage}
\begin{flushright}
BUHEP-93-2\\
hep-ph/9301222\\
January 8, 1993
\end{flushright}
\vspace{24mm}
\begin{center}
\Large{{\bf Triviality Bounds in
Two-Doublet Models}}
\end{center}
\vspace{5mm}
\begin{center}
Dimitris Kominis\footnote{e-mail address: kominis@budoe.bu.edu}
\ \  and \ \
R.~Sekhar Chivukula\footnote{e-mail address: sekhar@weyl.bu.edu}\\*[3.5mm]
{\normalsize\it Dept. of Physics, Boston University, 590 Commonwealth
Avenue,}\\
{\normalsize\it Boston, MA 02215}
\end{center}
\vspace{2cm}
\thispagestyle{empty}
\begin{abstract}
We examine
perturbatively the two-Higgs-doublet extension of the \SM\ in
the context of the suspected triviality of theories with fundamental scalars.
Requiring the model to define a consistent effective theory for scales below
a cutoff of $2\pi$ times the largest mass of the problem, as
motivated by lattice investigations of the one-Higgs-doublet model, we
obtain combined bounds for the parameters of the model. We find upper
limits of 470 GeV for the mass of the light $CP$--even neutral scalar
and 650--700 GeV for the other scalar masses.
\end{abstract}
\end{titlepage}
\newpage

\renewcommand{\thefootnote}{\arabic{footnote}}
\setcounter{footnote}{0}
\setcounter{page}{2}
\section{\bf Introduction}
In the Standard one-doublet Higgs Model
of electroweak interactions the scalar potential is
\be
V=\frac{1}{2}m_{0}^{2}\,\Phi^{\dag}\Phi+\frac{1}{4}\la_{0}\,
(\Phi^{\dag}\Phi)^{2}
\ee
\noindent
where $\Phi$ is a complex doublet and $m_{0}^{2}$, $\la_{0}$ are bare
parameters.
There are strong indications \cite{phi4,o(n)} that, in four dimensions and
in the limit of vanishing gauge and Yukawa couplings, this defines a
trivial field theory in the continuum limit. This means that for any
physically acceptable value of the bare coupling $\la_{0}$, the
renormalized self-coupling $\la_{R}$ is forced to lie in a narrow
range of values which shrinks to the point $\la_{R}=0$ at the limit
of infinite cutoff. Equivalently, a non-zero running coupling develops a
Landau pole at a finite momentum scale. Yukawa and gauge
couplings are not expected to alter this picture \cite{su2,shen,yuk}.
Consequently the Standard one-doublet Model
can only be accepted as an effective low energy
theory valid up to some finite cutoff $\La$. The value
of the renormalized coupling is thus allowed to be non-zero,
but is bounded from above.

This can
be illustrated perturbatively by integrating the one-loop
$\beta$-function for the scalar self-coupling. The result, ignoring
gauge and Yukawa couplings, is
\be
\frac{1}{\lambda(\mu)}=\frac{1}{\la(\La)}+\frac{3}{2\pi^{2}}\ln
\frac{\La}{\mu}
\ee
\noindent
Here $\la(\La)$ is the bare coupling and $\mu$ is some low energy
renormalization scale. Since $\la(\La) \geq 0$, it follows that 
\be
\la_{R}\equiv \la(\mu) \leq \,\frac{2\pi^{2}}{3}\,\frac{1}{\ln (\La/\mu)}
\ee
For a given cutoff $\La$, the mass $M_{H}$ of the Higgs boson is also
found to be bounded from above \cite{dn,o(n),su2,hasnag}. In
lowest-order perturbation theory
this is a consequence of the relation
\be
M_{H}^{2}=2\la_{R}\,v^{2}
\ee
where $v$ is the vacuum expectation value of the Higgs field.

Various physically motivated choices of $\La$ have been made leading to
different bounds on $M_{H}$ \cite{mai,cab,beg,cal,lind}. These bounds
generally increase with decreasing $\La$. For the effective theory
to make sense, the cutoff $\La$ must be at least of order $M_{H}$
\cite{dn}. This places an ``absolute'' upper bound on the mass of the
Higgs boson, which has been estimated \cite{o(n),hasnag,lind}
to be about 600--700 GeV, for small Yukawa and gauge couplings.

The purpose of this paper is to extend these considerations to models
with two Higgs doublets and derive bounds on the masses of the scalar
particles of these models. Our results are obtained using perturbative
arguments. We believe they convey the right qualitative picture and, in
the light of their agreement with other, non-perturbative approaches in
the case of the one-Higgs model, we expect they may also have some
quantitative validity.

In Section 2 we briefly review the two-doublet extension of the \SM. In
Section 3 we describe our calculation and in Section 4 we present and
discuss our results. For completeness, we list the renormalization group
equations for the couplings of the model in the appendix.

\vspace{0.65cm}
\section{The two-doublet model}
The scalar sector contains two electroweak doublets
$\Phi_{1},\;\Phi_{2}$, both with hypercharge $Y=1$. A discrete symmetry
must be imposed in order to eliminate flavor changing neutral currents
at tree level. The two-doublet models fall in two broad categories
according to the way this discrete symmetry is implemented \cite{sher}:
\be
\begin{array}{lcl}
\bullet\ \, {\rm Model\ I} & : & \Phi_{2}\rightarrow -\Phi_{2}
\mbox{\hspace{1cm} ;\hspace{1cm}}  d_{Ri}\rightarrow -d_{Ri}
\mbox{\hspace{51mm}}  \\*[2.5mm]
\bullet\ \, {\rm Model\ II} & : &
\Phi_{2}\rightarrow -\Phi_{2} \mbox{\hspace{28mm}}
\end{array}
\label{discrete}
\ee
($d_{Ri}$ ($i=1,2,3$) are the right-handed negatively charged quarks.)
The Lagrangian is
\bdm
{\cal L} ={\cal L}_{kin}+{\cal L}_{Y}-V
\edm
where ${\cal L}_{kin}$ contains all the covariant derivative terms, $V$ is the
scalar potential and
${\cal L}_{Y}$ contains the fermion-scalar interactions. The form of
the latter is the following:
\begin{itemize}
\item Model I
\be
{\cal L}_{Y}=g_{ij}^{(u)}\overline{\psi}_{Li}\Phi_{1}^{c}u_{Rj}+
g_{ij}^{(d)}\overline{\psi}_{Li}\Phi_{2}d_{Rj}\;+\;{\rm h.c.}\;+\;
{\rm leptons}
\ee
\item Model II
\be
{\cal L}_{Y}=g_{ij}^{(u)}\overline{\psi}_{Li}\Phi_{1}^{c}u_{Rj}+
g_{ij}^{(d)}\overline{\psi}_{Li}\Phi_{1}d_{Rj}\;+\;{\rm h.c.}\;+\;
{\rm leptons}
\ee
\end{itemize}
i.e. in Model I \,$\Phi_{1}$ gives mass to up-type quarks and $\Phi_{2}$  to
down-type quarks while in Model II only $\Phi_{1}$ couples to quarks.

The results we present were derived using Model II\@. Since the dominant
fermion effects are due to the top quark whose couplings are the same in
both models, no substantial changes are expected in Model I.

The scalar potential is
\begin{eqnarray}
V & = & \mu_{1}^{2}\,\Phi_{1}^{\dag}\Phi_{1}+
\mu_{2}^{2}\,\Phi_{2}^{\dag}\Phi_{2}
+\la_{1}\,(\Phi_{1}^{\dag}\Phi_{1})^{2}+\la_{2}(\Phi_{2}^{\dag}\Phi_{2})^{2}
+\la_{3}\,(\Phi_{1}^{\dag}\Phi_{1})(\Phi_{2}^{\dag}\Phi_{2}) \nonumber \\
  & & +\la_{4}\,(\Phi_{1}^{\dag}\Phi_{2})(\Phi_{2}^{\dag}\Phi_{1})
+\frac{1}{2}\la_{5}\,[(\Phi_{1}^{\dag}\Phi_{2})^{2}
+(\Phi_{2}^{\dag}\Phi_{1})^{2}] \label{potential}
\end{eqnarray}
\noindent
Note that by absorbing a phase in the definition of $\Phi_{2}$, we can
make $\la_{5}$ real and negative\footnote{This pushes all potential CP
violating effects into the Yukawa sector.}:
\be
\la_{5} \leq \, 0  \label{l5}
\ee
\noindent
The most interesting case arises when both doublets acquire non-zero
vacuum expectation values (vevs). To avoid spontaneous breakdown of the
electromagnetic $U(1)$, the vacuum expectation values must have the
following form:
\be
\langle\Phi_{1}\rangle=\frac{1}{\sqrt{2}} \left( \begin{array}{c}
0 \\ v_{1}
\end{array}  \right) \mbox{\hspace{1.1cm}  \hspace{1.1cm}}
\langle\Phi_{2}\rangle=\frac{1}{\sqrt{2}} \left( \begin{array}{c}
0 \\ v_{2}
\end{array}  \right)
\ee
where $v_{1}^{2}+v_{2}^{2}\equiv v^{2}=(246\ {\rm GeV})^{2}$.
The choice (\ref{l5}) ensures that $v_{1}$ and $v_{2}$ are
relatively real. ($v_{1}$ can be chosen to be real by an $SU(2)\times
U(1)$ rotation.) This configuration is indeed a minimum
of the tree level potential if
\be
\begin{array}{rlcrl}
\la_{1} & \geq \;0 & \mbox {\hspace{2cm}  \hspace{2cm}} & \la_{2} & \geq \;0
\\
\la_{4}+\la_{5} & \leq \;0 & \mbox {\hspace{2cm}  \hspace{2cm}} & 4\la_{1}
\la_{2} & \geq \; (\la_{3}+\la_{4}+\la_{5})^{2}
\end{array}
\label{minimum}
\ee

The spectrum of the scalar sector contains three Goldstone bosons, to be
eaten by the $W$'s and the $Z$; two neutral $CP$--even scalars, denoted
by $h$, $H$; one neutral $CP$--odd scalar $\zeta$; and two
charged scalars $G^{\pm}$. It is customary to introduce two angles
$\alpha$ and $\beta$:  $\beta$  $(0\!<\!\beta \!<\!\pi/2)$
rotates the $CP$--odd and the charged scalars into
their mass eigenstates while $\alpha$  $(-\pi/2 \leq \!\alpha \!<\!\pi/2)$
rotates the neutral scalars into
their mass eigenstates. The tree level expressions for the masses and
 angles are the following:
\newpage
\begin{eqnarray}
\tan \beta & = & \frac{v_{2}}{v_{1}} \label{mass1}\\
\sin \alpha & = & -({\rm sgn}\,C)\,\left[\frac{1}{2}
            \frac{\sqrt{(A-B)^{2}+4C^{2}}-(B-A)}{\sqrt{(A-B)^{2}+4C^{2}}}
            \right]^{1/2} \\
\cos \alpha & = & \left[\frac{1}{2}
            \frac{\sqrt{(A-B)^{2}+4C^{2}}+(B-A)}{\sqrt{(A-B)^{2}+4C^{2}}}
            \right]^{1/2} \\*[2mm]
M_{G^{\pm}}^{2} & = & -\frac{1}{2} (\la_{4}+\la_{5})\,v^{2} \\*[8mm]
M_{\zeta}^{2}\ \  & = & -\la_{5}\,v^{2} \\*[7.5mm]
\,M_{H,h}^{2} & = & \frac{1}{2}\left[A+B\pm \sqrt{(A-B)^{2}+4C^{2}}\right]
\end{eqnarray}
\noindent
where
\bdm
A=2\la_{1}\,v_{1}^{2} \mbox{\hspace{4mm};\hspace{4mm}}
B=2\la_{2}\,v_{2}^{2} \mbox{\hspace{4mm};\hspace{4mm}}
C=(\la_{3}+\la_{4}+\la_{5})\,v_{1}\,v_{2}
\edm

We emphasize that, as is the case in the one-doublet model,
all masses get their
scales from the vevs, with multiplicative factors that are functions of
the quartic couplings. If considerations of triviality put bounds on the
couplings (which they do), then these will automatically translate into
bounds for the masses.
The two-doublet models are described by 7 independent parameters which
can be taken to be $\alpha,\beta,M_{G^{\pm}},M_{\zeta},M_{h},M_{H}$ and
the top quark mass given by
\be
M_{t}=g_{t}\,v\cos \beta   \label{mt}
\ee
where $g_{t}$ is the top quark Yukawa coupling. The light quark and
lepton couplings are inessential to our analysis and we ignore them.
\vspace{0.65cm}

\section{Triviality and stability constraints}
We wish to determine when a given set of parameters
$\{\alpha,\beta,M_{G^{\pm}},M_{\zeta},M_{H},M_{h},M_{t}\}$ defines a valid,
consistent low energy effective theory. By `valid' we mean the
following: Suppose $\La$ is a finite cutoff scale beyond which new
phenomena appear. Any physical quantity calculated using the two-doublet
model as described in Section 2, will differ from its `true' value by
terms of order $p_{i}^{2}/\La^{2}$, $M_{j}^{2}/\La^{2}$ where $p_{i}$
are typical external momenta of the processes under consideration and
$M_{j}$ are the masses of the particles in the problem. We shall {\em
define} our theory to be a valid effective theory if all masses satisfy
\be
\frac{M_{j}}{\La} \leq \frac{1}{2\pi}  \label{2pi}
\ee
This convention corresponds to a Higgs
correlation length $M_{H}^{-1}=2$ (in lattice units), and
is widely used in lattice investigations of the problem
of triviality and Higgs mass bounds \cite{o(n)}.
(The external momenta $p_{i}$ should also satisfy a similar relation,
but this is irrelevant here.)
Thus, given a set of parameters, we define a cutoff
\be
\La=2\pi\,\max\left\{M_{G^{\pm}},M_{\zeta},M_{H},M_{h},M_{t},M_{Z}\right\}
\label{cutoff}
\ee
$M_{Z}$ being the $Z$-boson mass,
and require, for consistency of the theory, the following conditions to be
true:\\
(i) No coupling should develop a Landau pole at a scale less than
$\La$;\\
(ii) The effective potential should be stable for all field values less
than $\La$.\footnote{For field values greater than $\La$ the cutoff
effects are large and the renormalized effective potential is
meaningless. If a one-component Higgs-Yukawa system is well defined as a
bare theory, then it does not develop a vacuum instability \cite{shen}.
If this is the case in this model too, then the inequalities (\ref{stab})
are equivalent to the condition that the theory exists as a bare theory.}\\
The last requirement is satisfied if
\begin{eqnarray}
\la_{1}(\mu) & \geq & 0  \nonumber \\
\la_{2}(\mu) & \geq & 0  \label{stab} \\
\tilde{\la}(\mu) & \geq & -2\sqrt{\la_{1}(\mu)\,\la_{2}(\mu)}  \nonumber
\end{eqnarray}
for all $\mu \leq \La$, where
\be
\tilde{\la}(\mu)=\left\{ \begin{array}{lcl}
\la_{3}(\mu)+\la_{4}(\mu)+\la_{5}(\mu)  & \mbox{\hspace{5mm}{\rm
if}\hspace{5mm}}  &
\la_{4}(\mu)+\la_{5}(\mu) < 0 \\
\la_{3}(\mu)  & \mbox{\hspace{5mm}{\rm if}\hspace{5mm}} &
 \la_{4}(\mu)+\la_{5}(\mu) \geq 0
\end{array}
\right.
\ee

Our numerical procedure was the following: a set of parameters
$\{\alpha,\beta,M_{G^{\pm}},M_{\zeta},M_{H}$, $M_{h},M_{t}\}$
was chosen at random. By inverting the relations (\ref{mass1})--(\ref{mt})
the scalar and
Yukawa couplings were calculated. It was assumed that the tree-level
expressions (\ref{mass1})--(\ref{mt}) approximate best the physical values when
the
renormalization scale at which the couplings are evaluated is taken to
be
\be
\mu=\max\left\{M_{G^{\pm}},M_{\zeta},M_{H},M_{h},M_{t},M_{Z}\right\}
\ee
Note that (\ref{minimum}) are automatically satisfied if all masses are real.

The coupled renormalization group equations \cite{rg} for the scalar,
gauge and top Yukawa couplings
were evolved up to the scale defined by eq.~(\ref{cutoff}). (In practice,
$\La$ was taken to be at least 1 TeV which is the lowest scale at which
one would
expect new phenomena.) If any of the couplings became unbounded during
this evolution or if the stability constraints (\ref{stab}) were violated,
this set of parameters was rejected; otherwise it was accepted.
Subsequently a new set was chosen and the procedure repeated. In the
end, a large set of randomly generated `points' in parameter space
was accumulated. An envelope to these points represents the combined
bounds we are seeking.

\vspace{0.65cm}
\section{Results and discussion}
In Figures 2--8 we display projections of the allowed volume of
parameters on selected two-dimensional planes.
For comparison, in Fig. 1 we show
the bounds for the Standard one-doublet Model particles obtained using the same
method\footnote{Note the close agreement with the results of ref.\cite{o(n)}
where a relation equivalent to (\ref{2pi}) was used.}. The absolute bounds on
the masses of the scalar particles in the two-doublet model
are about 650--700 GeV (roughly the
same as the one-doublet model Higgs mass bound),
with the exception of the light
neutral scalar which is constrained to be lighter than about 470 GeV.
Upper bounds on the top quark are somewhat looser than in the Standard
one-doublet Model. We
estimate the numerical errors in the calculation of the bounds to be not
more than a few GeV, which is insignificant given the largely
qualitative nature of our computation. Experimental and other
theoretical bounds are not shown in these figures. The
upper limits on some splittings
among the scalar masses that arise from the precise measurement of the
electroweak $\rho$-parameter \cite{toussaint,hhg} are hardly more
stringent than our triviality bounds. Most other reported bounds are
lower bounds and do not interfere with our conclusions.

It is not possible to give a description of the exact shape of the
bounding surface in the parameter space. We will simply mention some
broad qualitative features: The bounds depend strongly on the angle
$\beta$; because of (\ref{mt}) the stability (lower) bounds become
stricter as $\beta$ becomes large
at fixed $M_{t}$. It is also found that for both small and large
$\beta$ the triviality bounds are stricter than they are for moderate
$\beta$; the precise way in
which this happens depends on the values of the other parameters. The
dependence on $\alpha$ is not as strong. Stability bounds on
the scalars are strictest when $\alpha$ takes values close to zero
(for a fixed top quark mass.) The bounds on $(M_{G^{\pm}},M_{\zeta})$
are largely insensitive to the values of $(M_{H},M_{h})$ for a large
range of these values, but shrink sharply outside that range ---and
vice-versa--- much like fig.~6 shows.

The angle $\alpha\!-\!\beta$ is of phenomenological significance since it
governs the couplings of the neutral scalars to the $W$'s and the $Z$.
We examined the bounds on the neutral scalar masses
as a function of $\cos^{2}(\alpha\!-\!\beta)$, projecting out all the
other parameters, and found no significant variation.

There is a way in which most of these bounds can be avoided, still
within the context of two-doublet models. A quadratic term
\begin{displaymath}
\mu_{3}^{2}\,\Phi_{1}^{\dag}\Phi_{2} \; + \; {\rm h.c.}
\end{displaymath}
can be added to the scalar potential (\ref{potential}).
This violates the discrete
symmetry (\ref{discrete}) but only softly, so that flavor changing neutral
currents still do not appear at tree level. In this case all scalar
particle masses but $M_{h}$ are increasing functions of $|\mu_{3}^{2}|$;
since  $\mu_{3}^{2}$ is not constrained from triviality considerations,
we can only impose bounds on $M_{h}$. As $|\mu_{3}^{2}|$ grows from
zero, we expect the bounds on $M_{G^{\pm}}$, $M_{\zeta}$ and $M_{H}$ to
become gradually weaker. For large $|\mu_{3}^{2}|$ there is a hierarchy
between the scales $M_{h}^{2}$ and $|\mu_{3}^{2}|$; the latter determines
the other scalar masses. Below $|\mu_{3}|$ the theory looks like the
one-Higgs model; insisting that the theory makes
sense as a {\em two-doublet model}
requires an effectively \SM\ quartic coupling to remain finite up to a
scale of order $2\pi|\mu_{3}|$ rather than $2\pi M_{h}$; hence we
expect much stricter bounds than those exhibited in Fig. 1. We have not
examined intermediate values of $|\mu_{3}^{2}|$ in more detail.

Bounds on the scalar particle masses from triviality considerations have
previously been reported in the literature. The authors of ref. \cite{fs,bw}
concentrate on very large cutoffs while in ref. \cite{maal} a different
definition of triviality, closely associated with perturbative
unitarity, is used. Our bounds are generally stricter than those imposed
by perturbative unitarity \cite{maal,hp}. The authors of
ref. \cite{joshi} adopt a similar,
but stricter, approach than ours and obtain a bound of 475 GeV for the
charged scalar mass $M_{G^{\pm}}$.

According to triviality constraints, the scalar sector of the one-Higgs
model is not allowed to become strongly interacting; even the heaviest possible
Higgs will be light enough to be detected as a relatively narrow
resonance at the SSC. We are currently investigating the implications of
the triviality and stability constraints on the phenomenology of
two-doublet models.

\vspace{0.9cm}
\noindent
{\Large\bf Acknowledgements}

\indent
We thank A.~Cohen, K.~Lane and Y.~Shen for reading the manuscript.
D.~K. would like to thank H.~Larralde and C.~Rebbi for suggestions in
the computational part of the work and V.~Koulovassilopoulos and Y.~Shen
for discussions.
R.S.C.\ acknowledges the support of an Alfred P. Sloan Foundation Fellowship,
an NSF Presidential Young Investigator Award, a DOE Outstanding Junior
Investigator Award, and a Superconducting Super Collider
National Fellowship from
the Texas National Research Laboratory Commission.  This work was
supported in part under NSF contract PHY-9057173 and DOE contract
DE-FG02-91ER40676, and by funds from the Texas National Research Laboratory
Commission under grant RGFY92B6.

\newpage
\vspace{0.9cm}
{\Large\bf Appendix}

\vspace{0.5cm}
In this appendix we include the coupled renormalization group equations for the
couplings of the two-doublet model \cite{rg}. The gauge couplings for the
$SU(3),SU(2)$ and $U(1)$ groups are $g_{c},\;g$ and $g^{\prime}$
respectively. For the other couplings we use the notation of the text.
We use the notation
\bdm
\cD \equiv 16\pi^{2}\,\mu\frac{d}{d\mu}
\edm

\begin{eqnarray*}
\cD g_{c}\: & = & -7g_{c}^{3} \\*[4mm]
\cD g\;\:   & = & -3g^{3}   \\*[4mm]
\cD g^{\prime}\: & = & 7g^{\prime 3}  \\*[4mm]
\cD g_{t}\: & = & g_{t}\left(
-\frac{17}{12}g^{\prime 2}-\frac{9}{4}g^{2}-8g_{c}^{2}+\frac{9}{2}g_{t}^{2}
\right) \\*[2.5mm]
\cD \la_{1} & = & 24\la_{1}^{2}+2\la_{3}^{2}+2\la_{3}\la_{4}
+\la_{4}^{2}+\la_{5}^{2}-3\la_{1}(3g^{2}+g^{\prime
2})+12\la_{1}g_{t}^{2}\\*[1mm]
  &  & +\frac{9}{8}g^{4}+\frac{3}{4}g^{2}g^{\prime 2}+\frac{3}{8}g^{\prime 4}
-6g_{t}^{4} \\*[4mm]
\cD \la_{2} & = & 24\la_{2}^{2}+2\la_{3}^{2}+2\la_{3}\la_{4}
+\la_{4}^{2}+\la_{5}^{2}-3\la_{2}(3g^{2}+g^{\prime
2})+\frac{9}{8}g^{4}+\frac{3}{4}g^{2}g^{\prime 2}
+\frac{3}{8}g^{\prime 4} \\*[4mm]
\cD \la_{3} & = &
4(\la_{1}+\la_{2})(3\la_{3}+\la_{4})+4\la_{3}^{2}+2\la_{4}^{2}+2\la_{5}^{2}
-3\la_{3}(3g^{2}+g^{\prime 2})+6\la_{3}g_{t}^{2}\\*[1mm]
  &   & +\frac{9}{4}g^{4}-
\frac{3}{2} g^{2}g^{\prime 2}+\frac{3}{4}g^{\prime 4} \\*[4mm]
\cD \la_{4} & = & 4\la_{4}(\la_{1}+\la_{2}+2\la_{3}+\la_{4})+8\la_{5}^{2}
-3\la_{4}(3g^{2}+g^{\prime 2})+6\la_{4}g_{t}^{2}+3g^{2}g^{\prime 2} \\*[4mm]
\cD \la_{5} & = & \la_{5}(4\la_{1}+4\la_{2}+8\la_{3}+12\la_{4}
-3(3g^{2}+g^{\prime 2})+6g_{t}^{2})
\end{eqnarray*}

\newpage

\noindent
{\Large\bf Figure captions}
\vspace{0.5cm}
\begin{enumerate}
\item
Triviality and stability bounds for the \SM\ Higgs and top quark masses
$M_{H}, M_{t}$. The allowed region is inside the curve.
\item
Triviality and stability bounds in the two-doublet model, for the heavy
neutral scalar $H$ and the top quark $t$. All other parameters are
projected on the $(M_{H},M_{t})$ plane: the region outside the curve is
excluded whatever the values of the parameters not shown on the graph.
Constraints from the weak interaction $\rho$-parameter suggest that
$M_{t} \simlt 250$ GeV \cite{rho}.
\item
Same as fig.~2, but projecting on the $(M_{h},M_{t})$ plane.
\item
Same as fig.~2, but projecting on the $(M_{G^{\pm}},M_{t})$ plane. A similar
graph is
obtained in the $(M_{\zeta},M_{t})$ plane, the bound on $M_{\zeta}$
being slightly higher than the one on $M_{G^{\pm}}$.
\item
Same as fig.~2, but projecting on the $(M_{H},M_{h})$ plane.
\item
Same as fig.~2, but projecting on the $(M_{H},M_{\zeta})$ plane. A similar plot
is
obtained for the $(M_{H},M_{G^{\pm}})$ plane, with the bounds on
$M_{G^{\pm}}$ slightly lower than those on $M_{\zeta}$.
\item
Same as fig.~2, but projecting on the $(M_{G^{\pm}},M_{\zeta})$ plane.
\item
Same as fig.~2, but projecting on the $(M_{h},M_{\zeta})$ plane; as in figures
4
and 6, the bounds on $M_{\zeta}$ are slightly higher than those on
$M_{G^{\pm}}$.
\end{enumerate}
\newpage
\thispagestyle{empty}
\begin{figure}
\includegraphics{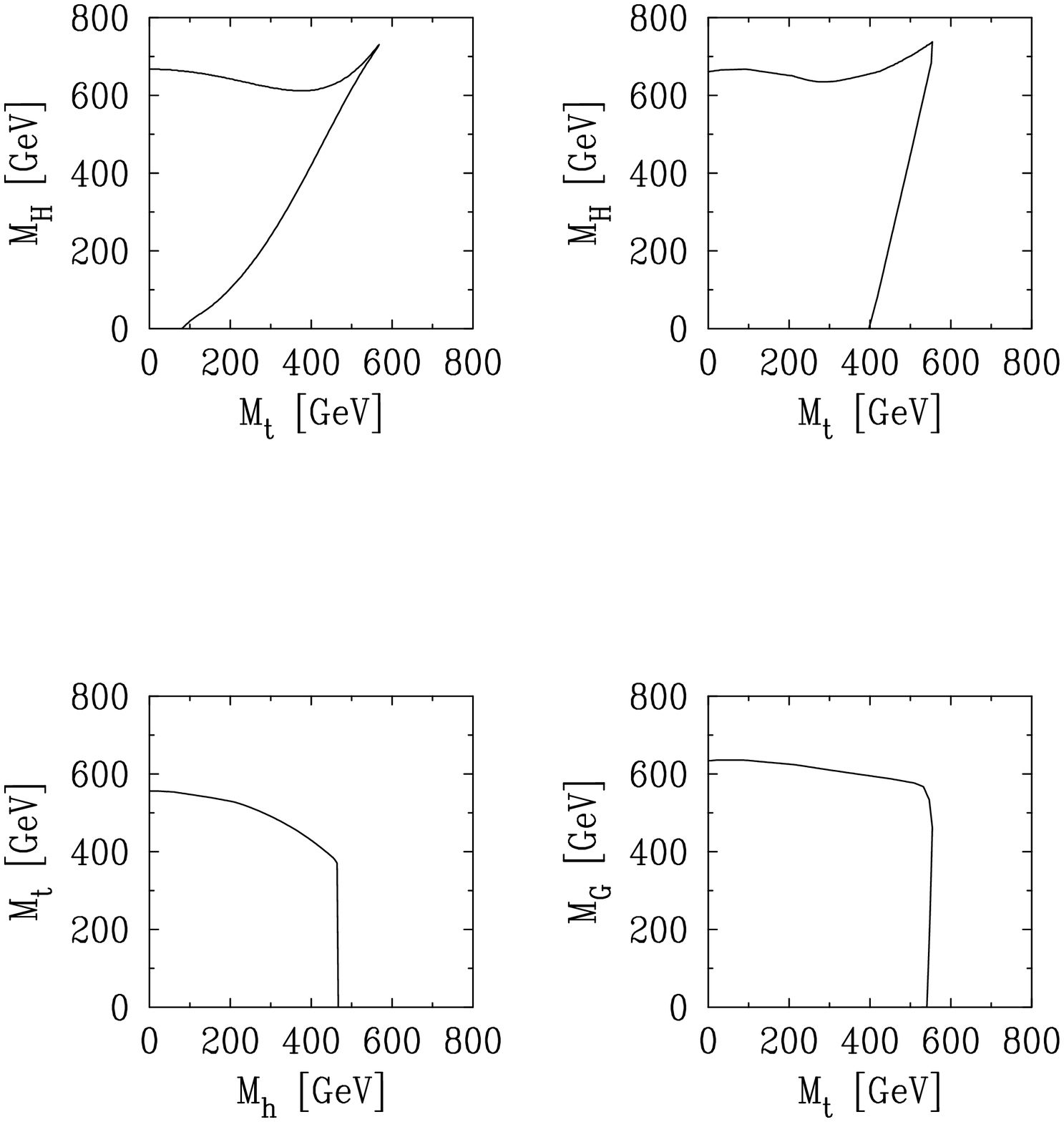}
\end{figure}
\mbox{     }

\vspace{7cm}
\hspace{2.1cm}\mbox{\large\bf {Figure 1}}\hspace{8.5cm}
\mbox{\large\bf {Figure 2}}

\vspace{12.2cm}
\hspace{2.1cm}\mbox{\large\bf {Figure 3}}\hspace{8.5cm}
\mbox{\large\bf {Figure 4}}
\newpage
\thispagestyle{empty}
\begin{figure}
\includegraphics{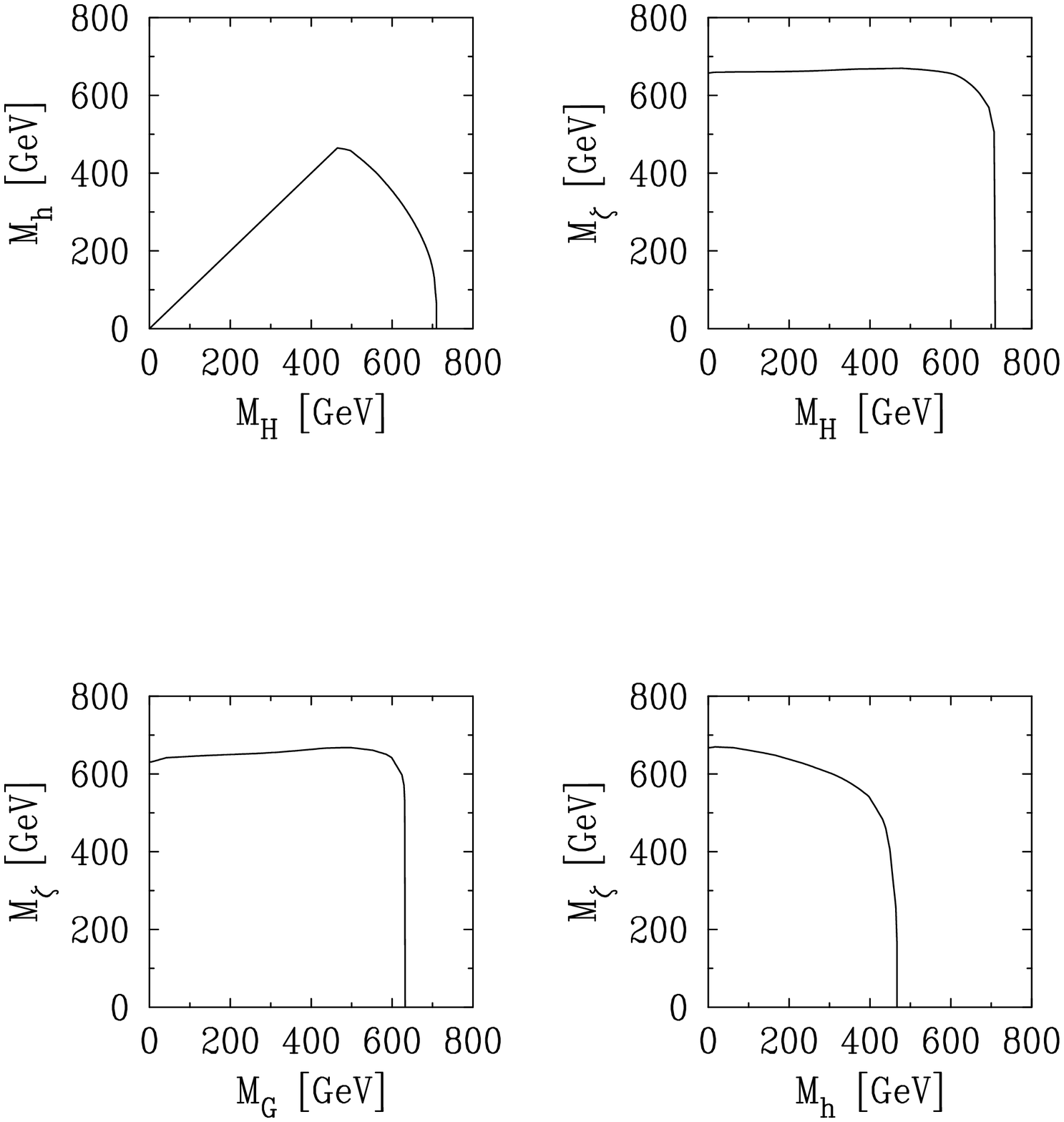}
\end{figure}
\mbox{     }

\vspace{7cm}
\hspace{2.1cm}\mbox{\large\bf {Figure 5}}\hspace{8.5cm}
\mbox{\large\bf {Figure 6}}

\vspace{12.2cm}
\hspace{2.1cm}\mbox{\large\bf {Figure 7}}\hspace{8.5cm}
\mbox{\large\bf {Figure 8}}


\newpage

\begin{thebibliography}{99}
\bibitem{phi4} K.~G.~Wilson, \pr {\bf B 4} (1971) 3184;\\
K.~G.~Wilson and J.~Kogut, \prp {\bf 12} (1974) 76;\\
G.~A.~Baker and J.~M.~Kincaid, \jsp {\bf 24} (1981) 469;\\
M.~Aizenmann, \prl {\bf 47} (1981) 1; \cmp {\bf 86} (1982) 1;\\
J.~Fr\"{o}hlich, \np{\bf 200} [FS4] (1982) 281;\\
A.~D.~Sokal, Ann. Inst. H. Poincar\'{e} {\bf A 37} (1982) 317;\\
C.~Arag\~{a}o de Carvalho, S.~Caracciolo and J.~Fr\"{o}hlich, \np{\bf
215} (1983) 209;\\
B.~Freedmann, P.~Smolensky and D.~Weingarten, \pl {\bf 113} (1982)
481;\\
I.~A.~Fox and I.~G.~Halliday, \pl {\bf 159} (1985) 148;\\
C.~B.~Lang, \pl {\bf 155} (1985) 399; \np{\bf 265} [FS15] (1986) 630;\\
I.~T.~Drummond, S.~Duane and R.~R.~Horgan, \np{\bf 280} [FS18] (1987)
25;\\
D.~J.~E.~Callaway and R.~Petronzio, \np{\bf 240} [FS12] (1984) 577;\\
K.~Gawedski and A.~Kupianen, \prl {\bf 54} (1985) 92;\\
M.~L\"{u}scher and P.~Weisz, \np{\bf 290} [FS20] (1987) 25;
\np{\bf 295} [FS21] (1988) 65.
\bibitem{o(n)} M.~L\"{u}scher and P.~Weisz, \np{\bf 318} (1989) 705;\\
J.~Kuti, L.~Lin and Y.~Shen, \prl {\bf 61} (1988) 678;\\
A.~Hasenfratz, K.~Jansen, C.~B.~Lang, T.~Neuhaus and H.~Yoneyama, \pl
{\bf 199} (1987) 531;\\
A.~Hasenfratz, K.~Jansen, J.~Jers\'{a}k, C.~B.~Lang, T.~Neuhaus and
H.~Yoneyama, \np{\bf 317} (1989) 81;\\
G.~Bhanot, K.~Bitar, U.~M.~Heller and H.~Neuberger, \np{\bf 353} (1991) 551.
\bibitem{su2} A.~Hasenfratz and P.~Hasenfratz, \pr {\bf D 34} (1986)
3160;\\
I.~Montvay, \np{\bf 293} (1987) 479;\\
W.~Langguth, I.~Montvay and P.~Weisz, \np{\bf 277} (1986) 11; \zp {\bf
36} (1987) 725;\\
A.~Hasenfratz and T.~Neuhaus, \np{\bf 297} (1988) 205.
\bibitem{shen} Y.~Shen, Nucl.\ Phys.\ (Proc.\ Suppl.) {\bf B20} (1991)
613.
\bibitem{yuk} R.~Shrock, Nucl.\ Phys.\ (Proc.\ Suppl.) {\bf B20} (1991)
585;\\
W.~Bock, Nucl.\ Phys.\ (Proc.\ Suppl.) {\bf B20} (1991) 559;\\
K.~Jansen Nucl.\ Phys.\ (Proc.\ Suppl.) {\bf B20} (1991) 564.
\bibitem{dn} R.~Dashen and H.~Neuberger, \prl {\bf 50} (1983) 1897.
\bibitem{hasnag} P.~Hasenfratz and J.~Nager, \zp {\bf 37} (1988) 477.
\bibitem{mai} L.~Maiani, G.~Parisi and R.~Petronzio, \np{\bf 136} (1978)
115.
\bibitem{cab} N.~Cabibbo, L.~Maiani, G.~Parisi and R.~Petronzio, \np{\bf
158} (1979) 295.
\bibitem{beg} M.~A.~B.~B\'{e}g, C.~Panagiotakopoulos and A.~Sirlin, \prl
{\bf 52} (1984) 883.
\bibitem{cal} D.~Callaway, \np{\bf 233} (1984) 189.
\bibitem{lind} M.~Lindner, \zp {\bf 31} (1986) 295.
\bibitem{sher} M.~Sher, \prp {\bf 179} (1989) 273.
\bibitem{rg} T.~Cheng, E.~Eichten and L.-F.~Li, \pr {\bf D 9} (1974)
2259;\\
K.~Inoue, A.~Kakuto and Y.~Nakano, \ptp {\bf 63} (1980) 234.
\bibitem{toussaint} D.~Toussaint, \pr {\bf D 18} (1978) 1626.
\bibitem{hhg} J.~F.~Gunion, H.~E.~Haber, G.~L.~Kane and S.~Dawson, The
Higgs hunter's guide, (Addison-Wesley, Reading, MA, 1990).
\bibitem{fs} R.~Flores and M.~Sher, \ap {\bf 148} (1983) 95.
\bibitem{bw} A.~Bovier and D.~Wyler, \pl {\bf 154} (1985) 43.
\bibitem{maal} J.~Maalampi, J.~Sirkka and I.~Vilja, \pl {\bf 265} (1991)
371.
\bibitem{hp} H.~H\"{u}ffel and G.~P\'{o}csik, \zp {\bf 8} (1981) 13.
\bibitem{joshi} A.~J.~Davies and G.~C.~Joshi, \prl {\bf 58} (1987) 1919.
\bibitem{rho} ALEPH Collab., D.~Decamp at al., \zp {\bf 53} (1992) 1.
\end{thebibliography}
\end{document}